\documentclass[%
superscriptaddress,
twocolumn,
amsmath,
amssymb,
 aps,
pra,
floatfix,
]{revtex4-2}

\usepackage{graphicx}
\usepackage[colorlinks]{hyperref}
\hypersetup{%
	plainpages=true,
	breaklinks=true,
	hypertexnames=false,
	pageanchor=true,
	colorlinks=true,
	linkcolor={blue},
	citecolor={red},
	urlcolor={blue},
	anchorcolor={black}
}

\usepackage[dvipsnames]{xcolor}
\usepackage[normalem]{ulem}
\usepackage{amsthm}
\usepackage{mathtools}

\newcommand{\bs}[1]{\boldsymbol{#1}} 
\newcommand{\rmm}{{\rm m}} 
\newcommand{\rmeff}{{\rm eff}} 

\begin{document}
\title{
Bound polariton states in the Dicke-Ising model
}

\author{Juan Román-Roche}
\affiliation {Instituto de Nanociencia y Materiales de Aragón (INMA), CSIC-Universidad de Zaragoza, Zaragoza 50009, Spain}
\affiliation{Departamento de Física de la Materia Condensada, Universidad de Zaragoza, Zaragoza 50009, Spain}

\author{Álvaro Gómez-León}
\affiliation {Institute of Fundamental Physics IFF-CSIC, Calle Serrano 113b, 28006 Madrid, Spain}

\author{Fernando Luis}
\affiliation {Instituto de Nanociencia y Materiales de Aragón (INMA), CSIC-Universidad de Zaragoza, Zaragoza 50009, Spain}
\affiliation{Departamento de Física de la Materia Condensada, Universidad de Zaragoza, Zaragoza 50009, Spain}

\author{David Zueco}
\affiliation {Instituto de Nanociencia y Materiales de Aragón (INMA), CSIC-Universidad de Zaragoza, Zaragoza 50009, Spain}
\affiliation{Departamento de Física de la Materia Condensada, Universidad de Zaragoza, Zaragoza 50009, Spain}
  
\date{\today}

\begin{abstract}
We present a study of hybrid light-matter excitations in cavity QED materials using the Dicke-Ising model as a theoretical framework.
Leveraging linear response theory, we derive the exact excitations of the system in the thermodynamic limit. 
Our results demonstrate that the cavity can localize spin excitations, leading to the formation of bound polaritons, where the cavity
acts as an impurity of the
two-excitation band, localizing spin-wave pairs around
single-spin domains.
We derive the condition for the existence of these bound states and discuss its satisfiability in different regimes.
Finally, we show that these effects persist in finite systems using exact-diagonalization calculations.

\end{abstract} 

\maketitle

\section{Introduction}

The control of quantum matter with quantum light is a common pursuit in quantum optics. Initially, the focus was on minimalistic matter such as single atoms and molecules. Due to the weak light-matter coupling, it was realized that photons need to be confined in cavities, giving rise to cavity quantum electrodynamics (cQED) \cite{haroche2012nobel, wineland2012nobel}. This field has since evolved to consider more complex forms of matter as well. First, using the cavity as a probe for materials in cavity-enhanced spectroscopy, and more recently, to push the boundaries of light-matter interaction to explore whether quantum light, either a few photons or vacuum states, can alter the properties of matter \cite{schlawin2022cavity, garciavidal2021manipulating, bloch2022strongly}. Seminal experimental demonstrations modifying and controlling conductivity \cite{paravicinibagliani2018magnetotransport, appugliese2022breakdown}, magnetism  \cite{thomas2021large} and the metal-to-insulator transition \cite{jarc2023cavity} led to envisioning novel phenomenology emerging from the hybridization of light and matter, such as modifications of chemical reactions \cite{feist2017polaritonic, nagarajan2021chemistry, fregoni2022theoretical}, changes in the critical temperature in superconductivity \cite{schlawin2019cavitymediated, sentef2018cavity}, or alterations in magnetism \cite{romanroche2021photon, masuki2024cavity}, topology \cite{dmytruk2022controlling, downing2019topological, perezgonzalez2022topology, perezgonzalez2023lightmatter,dmytruk2023hybrid,perezgonzalez2023manybody}, ferroelectricity \cite{schuler2020thevacua, ashida2020quantum, shin2022simulating} and transport in excitonic \cite{feist2015extraordinary}, molecular \cite{sandik2024cavityenhanced} and disordered electronic systems \cite{ciuti2021cavitymediated, arwas2023quantum, svintsov2024onedimensional}.
\begin{figure}[t]
    \centering
    \includegraphics[width=\columnwidth]{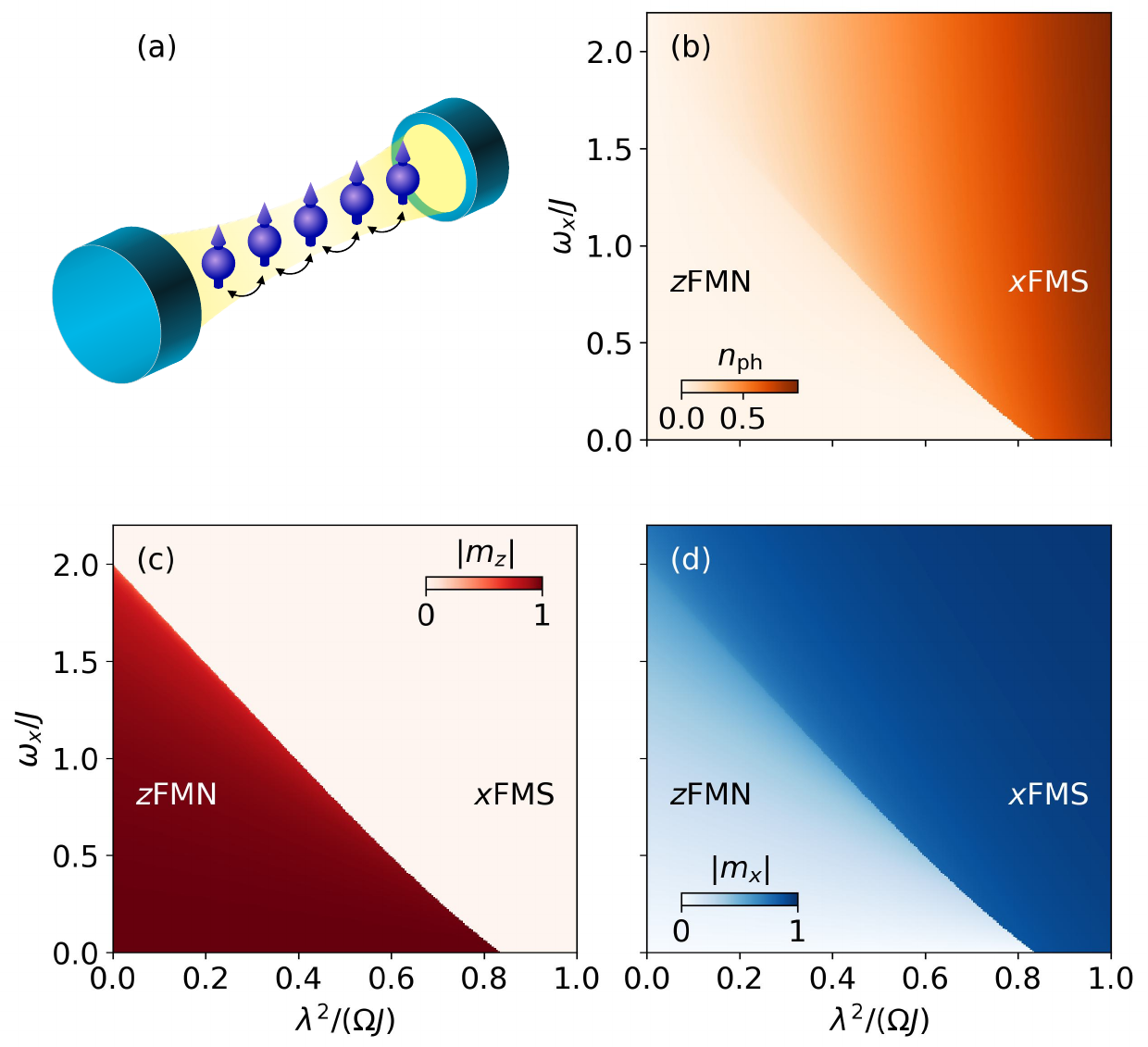}
    \caption{Sketch (a) and phase diagram (b, c, d) of the Dicke-Ising model in the $(\lambda^2/\Omega, \omega_x)$ plane. 
    (b) Number of photons, $n_{\rm ph}$, which is an order parameter of the $x$-ferromagnetic superradiant ($x$FMS) phase for $\omega_x = 0$.
    (c) Longitudinal magnetization, $m_z$, which is the order parameter of the $z$-ferromagnetic normal ($z$FMN) phase.
    (d) Transverse magnetization, $m_x$, which is another order parameter of the $x$FMS phase for $\omega_x = 0$.}
    \label{fig:dickeisingphasediagram}
\end{figure}

Matter alterations can be underlain by the modification of the ground (thermal) state, and/or from changes to the excitation spectrum. 
The mixing of two near-resonant energy levels gives rise to polaritons, hybrid states of light and matter that exhibit properties of both constituents \cite{basov2020polariton}.
Here, we focus on a key scenario of excitation hybridization: when a continuum couples to a discrete level it can give rise to new discrete levels outside of the continuum, known as bound states.
In waveguide QED, the role of the continuum is played by the electromagnetic modes of the waveguide and a coupled emitter provides the discrete energy level.
The resulting bound state is spatially localized around the emitter.
These bound states have attracted significant attention due to their ability to control light emission, such as inhibiting or enhancing spontaneous emission \cite{Khalfin1958,Bykov1975,Fonda1978,Onley1992,gaveau1995limited,Garmon2009,Garmon2013,Garmon2013b,Lombardo2014,sanchez2017b}. 
They can also mediate long-range interactions between emitters \cite{GonzalezTudela2015,Douglas2015,shi2016,calajo2016,GonzalezTudela2017,GonzalezTudela2017b,GonzalezTudela2018,GonzalezTudela2018b,GonzalezTudela2018d,shi2018c,bello2019,sanchez2019b,romanroche2020bound}.
In cavity QED materials, light and matter exchange their roles with respect to waveguide QED, as it is common to consider a macroscopic material that hosts a continuum of energy levels, in the form of bands, coupled to a single cavity mode \cite{lenk2020collective, debernardis2022magneticfieldinduced, dmytruk2022controlling, bacciconi2023firstorder, vlasiuk2023cavityinduced}.

In this paper, we discuss the emergence of bound polaritons in cavity QED materials.  
These are localized bound states arising from the hybridization of the material energy band with the cavity mode.
For this purpose, we employ the Dicke-Ising model, \emph{i.e.}, a spin-$1/2$ Ising chain coupled transversally to the quantum field fluctuations of the cavity, see Fig. \ref{fig:dickeisingphasediagram}(a). It generalizes the Dicke model by introducing intrinsic (Ising) interactions among the two-level systems, and it extends the Ising model by considering a quantum transverse field.
While it serves as a toy model for a magnetic material coupled to a cavity, it can also be experimentally realized with an array of superconducting qubits coupled to a one-dimensional transmission-line resonator \cite{zhang2014quantum}.
Crucially, it is exactly solvable in the thermodynamic limit ($N \to \infty$ with $N$ the number of spins)  and its phase diagram is well known \cite{lee2004firstorder, gammelmark2011phase}. 
Additionally, using a linear response theory developed by us for cavity QED materials, the excitations can be obtained exactly \cite{romanroche2024cavity}. 
This allows us to determine the conditions for the existence of bound polaritons and establish a connection with the bound states in waveguide QED.
We show the formation of localized polaritonic bound states hybridizing spin-wave pairs and the cavity photon near the band edges. 
The reason for this is that the model, through a Jordan-Wigner transformation, can be mapped onto an impurity model, more specifically, a boson localized within the \emph{real} space of a continuum of fermions.

The rest of the paper is organized as follows. In Section \ref{sec:model}, the 
light-matter model is presented. Section, \ref{sec:solution}, is the 
main part of our work, where we solve the Dicke-Ising model, including its 
equilibrium and linear response. Additionally, we prove the existence of bound 
polariton states. We also perform exact diagonalization calculations 
for finite systems. Tautologically, we conclude with the conclusions. The
microscopic theory of the light-matter Hamiltonian is presented in appendix 
\ref{app:microscopic}, and the calculation of the dressed material response is 
presented in appendix \ref{app:responseising}.

\section{Model}
\label{sec:model}

The light-matter Hamiltonian for a magnetic material coupled to a uniform cavity mode reads
\begin{equation}
    H = H_{\rm m} - \frac{\lambda}{\sqrt{N}} \sum_j^N \frac{1}{\mu_B} \bs m_j \cdot \bs u \left( a + a^\dagger \right) + \Omega a^\dagger a \,,
    \label{eq:Hzeeman}
\end{equation}
with $a$, $a^\dagger$ bosonic annihilation and creation operators, $[a, a^\dagger]=1$, and $\bs m_j$ the magnetic dipole operators of the material [See App. \ref{app:microscopic} for a derivation].
Here $H_\rmm$ is the Hamiltonian of the bare magnetic material and $\lambda/\sqrt{N} = \mu_{\rm B} B$ is the Zeeman coupling to the magnetic field of the cavity, $\bs B = B \bs u$. Importantly, we consider the material in the thermodynamic limit for the number of magnetic dipoles, $N \to \infty$. The cavity field intensity depends on the inverse square root of the mode volume $B \sim 1/\sqrt{V}$. To ensure a well-defined thermodynamic limit, we assume a finite density of dipoles in the cavity, $\rho = N/V = {\rm cst.} \,$. Accordingly, we find that $B \sim 1/\sqrt{N}$ for $N \to \infty$.

We will consider a toy model of a magnetic material, the spin-$1/2$ Ising chain in transverse field, with transverse coupling to the cavity, such that
\begin{equation}
    H_{\rm m} = \frac{\omega_x}{2} \sum_j^N\sigma_j^x - J \sum_j^N\sigma_j^z \sigma_{j+1}^z
    \label{eq:Hising}
\end{equation}
and 
\begin{equation}
    \frac{1}{\mu_B} \bs m_j \cdot \bs u = \sigma_j^x \,,
    \label{eq:couplingising}
\end{equation}
with $\sigma_j^\alpha$ the Pauli matrices, $[\sigma_i^\alpha, \sigma_j^\beta] = \delta_{ij} 2 \epsilon_{\alpha \beta \gamma} \sigma_i^\gamma$ and $\epsilon_{\alpha \beta \gamma}$ the Levi-Civita symbol.
The full light-matter model is termed the (transverse) Dicke-Ising model \cite{lee2004firstorder, gammelmark2011phase, zhang2014quantum, cortese2017polariton, rohn2020ising, schellengerger2024almost, langheld2024quantum}. For vanishing transverse field the model has a $\mathbb Z_2 \times \mathbb Z_2$ symmetry. The first symmetry corresponds to a spin-flip, $\sigma_j^z \to - \sigma_j^z$, and in the bare Ising model it is spontaneously broken in a second-order phase transition from a paramagnetic to a ferromagnetic phase. The second symmetry corresponds to a simultaneous cavity-field and spin flip, $a \to -a$ and $\sigma_j^x \to -\sigma_j^x$, and in the bare Dicke model it is spontaneously broken in a second-order phase transition from a paramagnetic normal to a ferromagnetic superradiant phase. Their combination gives rise to a first-order phase transition in the Dicke-Ising model between two symmetry-broken phases: an $x$-ferromagnetic superradiant ($x$FMS) phase for large $g^2/\Omega$ and a $z$-ferromagnetic normal ($z$FMN) phase for large $J$. A non-zero classical transverse field breaks the Dicke symmetry but the model still features a first-order phase transition between the $x$FMS phase where the order direction is fixed by the classical field to a symmetry-broken $z$FMN phase.
Alternatively, the Ising chain with longitudinal coupling to the cavity has been studied in Ref. \cite{mckenzie2022theory}.
The model defined by Eqs. \eqref{eq:Hzeeman} and \eqref{eq:Hising} is sketched in Fig. \ref{fig:dickeisingphasediagram}(a).

\section{Exact solution in the thermodynamic limit}
\label{sec:solution}

Following \cite{romanroche2024cavity}, the equilibrium and linear response properties of
model \eqref{eq:Hzeeman} can be computed exactly in the thermodynamic limit, $N \to \infty$.
This is essentially because the cavity mediates collective all-to-all interactions between the spins, which can be treated exactly with a mean-field approach.

\subsection{Ground state phase diagram}

The equilibrium properties are obtained by solving the mean-field effective Hamiltonian \cite{romanroche2022effective}
\begin{equation}
    H_{\rm{eff}}^{\rm{MF}} = \frac{\tilde \omega_x}{2} \sum_j^N\sigma_j^x - J \sum_j^N\sigma_j^z \sigma_{j+1}^z + \frac{N \lambda^2}{\Omega} m_x^2 \,,
    \label{eq:HeffMFising}
\end{equation}
with $\tilde \omega_x = \omega_x - 4 \lambda^2/\Omega m_x$ and $m_x = N^{-1} \sum_j^N\langle \sigma_j^x \rangle$, variationally with respect to $m_x$. Then, photonic observables can be computed from the relation $\langle a \rangle = \sqrt{N} \lambda /\Omega m_x$.
Equation \eqref{eq:HeffMFising} corresponds to the Ising chain in a transverse field.
The transverse field is a combination of the external field and the cavity-induced mean field. 
It is now clear that adding a longitudinal field would make $H_{\rm{eff}}^{\rm{MF}}$ analytically intractable, as the resulting mean-field effective Hamiltonian would correspond to the Ising model with both transverse and longitudinal fields. 
In the thermodynamic limit, $N \to \infty$, the ground-state energy per spin is given by \cite[Chap. 10]{sachdevquantum}
\begin{equation}
    \tilde e_0(m_x) = \frac{\lambda^2}{\Omega} m_x^2 - \frac{1}{2} \int_{-\pi}^{\pi} \frac{dk}{2 \pi} \tilde \epsilon_k \,,
\end{equation}
with
\begin{equation}
    \tilde \epsilon_k = \sqrt{(2J)^2 + \tilde \omega_x^2 - 4 J \tilde \omega_x \cos k} \,.
    \label{eq:isingdisp}
\end{equation}
Solving variationally allows us to compute the equilibrium value of $m_x$ numerically and subsequently the longitudinal magnetization as \cite{pfeuty1970the}
\begin{equation}
    m_z = \begin{cases}
        \left(1 - \left(\frac{\tilde \omega_x}{2J}\right)^2\right)^{\frac{1}{8}} & {\rm if} \quad 0 \leq \frac{\tilde \omega_x}{2J} \leq 1 \,, \\
        0 & {\rm if} \quad  1 < \frac{\tilde \omega_x}{2J} \,.
    \end{cases}
    \label{eq:mz}
\end{equation}
and the number of photons per spin as
\begin{equation}
    n_{\rm ph} = \frac{\lambda^2}{\Omega^2} m_x^2 \,.
    \label{eq:nph}
\end{equation}

The zero-temperature phase diagram is presented in Fig. \ref{fig:dickeisingphasediagram}.
The transverse magnetization, $m_z$, acts as an order parameter for the $z$FMN phase.
For $\omega_x = 0$ the longitudinal magnetization, $m_x$, and the photon number, $n_{\rm ph}$, act as order parameters for the $x$FMS phase.
An analysis of $e_0(m_x)$ in this case reveals that the system undergoes a first-order phase transition at $\lambda^2/(\Omega J) \approx 0.837$.
In the opposite case of vanishing light-matter coupling, $\lambda = 0$, the Ising chain in tranverse field is known to undergo a second order phase transition at $\omega_x/J = 2$.
In previous solutions of the Dicke-Ising model the transverse field is set in a direction perpendicular to both the intrinsic interaction and the cavity field, which would be the $y$ direction in our case \cite{lee2004firstorder,gammelmark2011phase}.
Although this difference is subtle, in implies that $m_x$ is always an order parameter of the $x$FMS phase.
Then, a Landau analysis of the ground-state energy in terms of its series expansion in powers of $m_x$ reveals the existence of a tricritical point splitting the critical line into a regime of second-order criticality for large $\omega_x$ and small $\lambda$ and a regime of first-order criticality for small $\omega_x$ and large $\lambda$ \cite{gammelmark2011phase}.
In the present case, where the classical transverse field and the cavity fields are aligned, the Landau analysis is not possible.
Nevertheless, a visual inspection of the landscape of energy minima of $e_0(m_x)$ reveals the existence of a tricritical point at $\lambda^2/(\Omega J) \approx 0.225$ and $\omega_x/J = 1.427$.

\subsection{Linear response theory}

\begin{figure}
    \centering
    \includegraphics[width=\columnwidth]{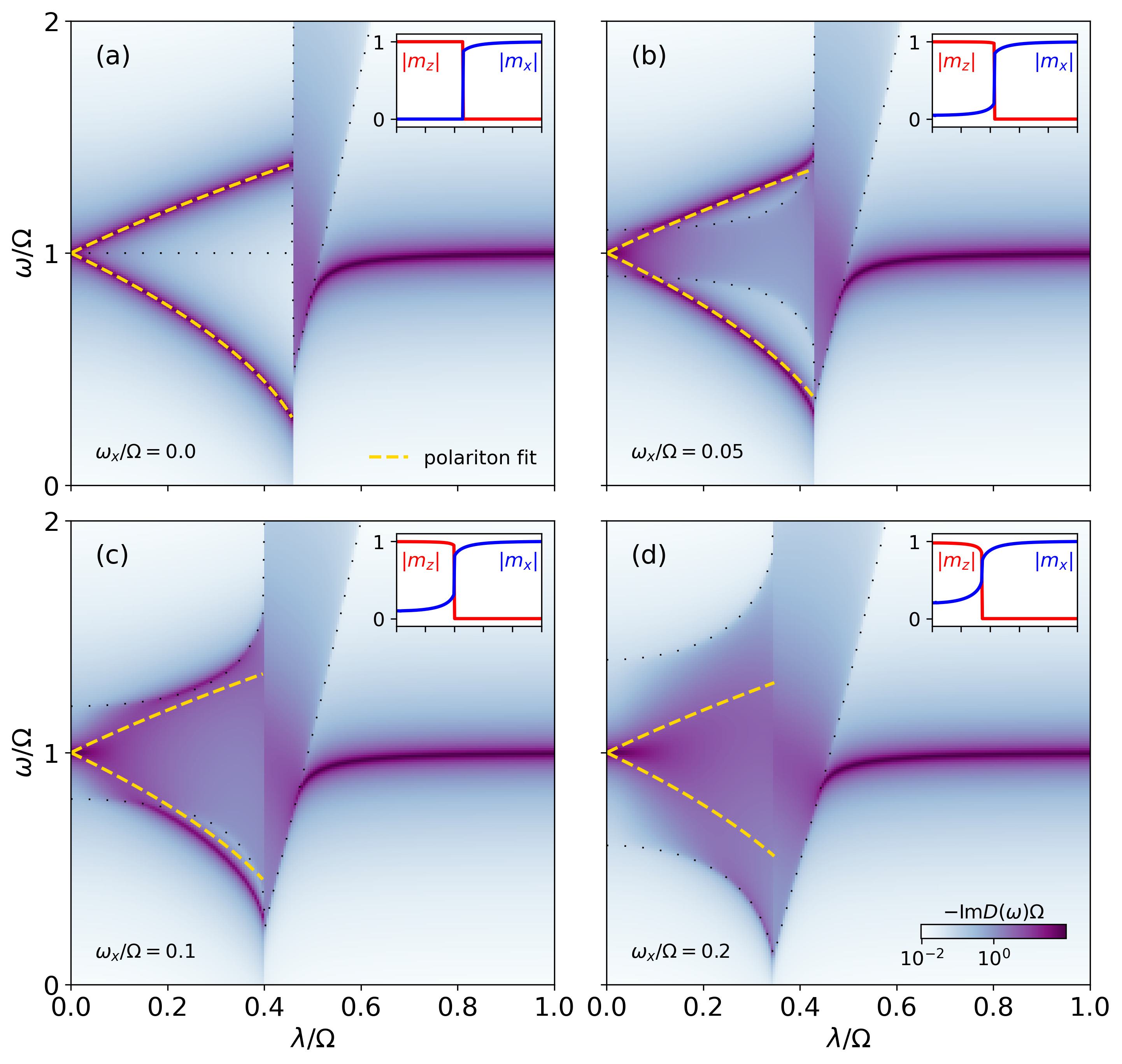}
    \caption{Cavity response, $D$, of the Dicke model as a function of the collective coupling, $\lambda$, for different values of the classical transverse field, $\omega_x$. 
    The yellow dashed lines correspond to a fit of the polaritons with a two-oscillator model [See Eq. \eqref{eq:2osci}]. 
    The top right insets show the magnetization. The dotted lines mark the edges of the band of the mean-field effective Hamiltonian \eqref{eq:HeffMFising}. 
    The Ising interaction is set to $4J = \Omega$.}
    \label{fig:dickeisingboundstates}
\end{figure}

The response functions of the hybrid system are given by retarded Green functions \cite[Chap. 7]{altland2010condensed}. The retarded Green function for operators $A$ and $B$ is defined as
\begin{equation}
    G^{\rm r}_{A, B}(t, t') = -i\theta(t - t') \langle [A(t), B(t')]\rangle \,.
    \label{eq:retardedgreenfunction}
\end{equation}
We will be particularly concerned with the photonic propagator
\begin{equation}
    D(t) = G^{\rm r}_{a, a^\dagger}(t, 0) \,,
\end{equation}
and the matter response function
\begin{equation}
    \chi(t) = -\frac{1}{N} G^{\rm r}_{C_x, C_x}(t, 0) \,,
\end{equation}
for the coupling operator $C_x = \sum_j^N\sigma_j^x$.
In the thermodynamic limit, these are given by
\begin{equation}
    D(\omega) = D_0(\omega) - \lambda^2 D_0(\omega) \chi(\omega) D_0(\omega) \,.
    \label{eq:relpropphtonmatter}
\end{equation}
and
\begin{equation}
    \chi(\omega) = \frac{\tilde \chi_{0}(\omega)}{1 + V_{\rm ind}(\omega) \tilde \chi_{0}(\omega)} 
    \label{eq:chixaidentity}
\end{equation}
with 
\begin{equation}
    V_{\rm ind}(\omega) = \frac{2 \lambda^2}{\Omega} \frac{\Omega^2}{(\omega + i 0)^2 - \Omega^2} \,,
\end{equation}
Here $D_0$ is the bare photonic propagator
\begin{equation}
    D_0(\omega) = \frac{1}{\omega + i0 - \Omega} \,,
\end{equation}
and $\tilde \chi_{0}$ is the matter response function for the mean-field effective matter Hamiltonian of Eq. \eqref{eq:HeffMFising}, i.e.
\begin{equation}
    \tilde \chi_{0}(t) = \frac{i}{N}  \theta(t) \langle [C_x(t), C_x(0)]\rangle_{H_\rmeff^{\rm MF}} \,.
\end{equation}
Note that in the cases where $m_x = 0$, $H_\rmeff^{\rm MF} = H_\rmm$ and thus $\tilde \chi_{0} = \chi_{0}$ is the bare matter response function.

At zero temperature and in the continuum limit, $\tilde \chi_{0}$ is given by (See App. \ref{app:responseising} for details)
\begin{equation}
    \tilde \chi_{0}(\omega) = -32J^2 \int_{-\pi}^{\pi} \frac{dk}{2 \pi} \frac{\sin^2 k}{\tilde \epsilon_k ((\omega + i0)^2 - 4 \tilde \epsilon_k^2)} \,.
    \label{eq:chi0ising}
\end{equation}
Interestingly, we find that $\tilde \chi_{0}$ has poles at $\omega = 2 \tilde \epsilon_k$. 
This stems from the fact that the coupling operator, $C_x$, creates and destroys excitations in pairs of opposite momentum
\begin{equation}
\begin{multlined}
    C_x = N - 2 \sum_k \bigl(\tilde v_k^2 + (\tilde u_k^2 - \tilde v_k^2) \tilde \gamma_k^\dagger \tilde \gamma_k \\
    + i \tilde u_k \tilde v_k (\tilde \gamma_k^\dagger \tilde \gamma_{-k}^\dagger - \tilde \gamma_{-k} \tilde \gamma_k) \bigr) \,,
\end{multlined}
\label{eq:Cxbogoliubovising}
\end{equation}
where $\tilde \gamma_k$ and $\tilde \gamma_k^\dagger$ are the annihilation and creation operators of the Bogoliuvov fermions that constitute the elementary excitations of the effective Ising model \eqref{eq:HeffMFising} after a Jordan-Wigner fermionization \cite[Chap. 10]{sachdevquantum}. Here $\tilde u_k = \cos(\tilde \theta_k /2)$ and $\tilde v_k=\sin(\tilde \theta_k /2)$ are the Bogoliubov coefficients, with
\begin{equation}
    \tan \tilde \theta_k = \frac{\sin k}{\frac{\tilde \omega_x}{2J} - \cos k} \,.
\end{equation}
Thus, $\chi$ and $D$ will reflect how the zero-momentum sector of the two-excitation band (a double-energy replica of the single-excitation band) of the Ising model hybridizes with the cavity photon. 
The fact that excitations are created in pairs of opposite momenta allows one to probe the full band, despite the collective nature of $C_x$.  

This feature brings novel phenomenology that we summarize in Fig. \ref{fig:dickeisingboundstates}, where the cavity response \eqref{eq:relpropphtonmatter} is plotted for different scenarios. 
In all panels, we set $4J = \Omega$, such that the two-excitation band is in resonance with the cavity frequency. 
Figure \ref{fig:dickeisingboundstates}(a) shows the case of vanishing classical field, $\omega_x=0$. 
In this case, the model is non-dispersive in the normal phase, as the only source of transverse field is the effective mean field. 
Instead of a band, the model has a collection of degenerate excitations with energy $2J$ that are linear combinations of domain walls \cite{prelovsek1981quantum}. 
Accordingly, the zero-momentum sector of the two-excitation band is a degenerate collection of the double excitations that correspond to single-spin flips. 
This is a typical situation where the cavity is coupled to a collective mode hybridizing with the cavity photon, forming polaritons whose energy can be fitted by a model of two coupled quantum harmonic oscillators of frequencies $\Omega$ and $4J$:
\begin{equation}
    2 \Omega_\pm^2 = 4J^2 + \Omega^2 \pm \sqrt{\left(4J^2 - \Omega^2\right)^2 + 32 \lambda^2 J \Omega} \,.
    \label{eq:2osci}
\end{equation}
At the first-order phase transition, the effective mean field acquires a non-zero value, opening the band. 
The lower polariton hardens to become the cavity photon in the deep superradiant regime \cite{ashida2021cavity}.

Figures \ref{fig:dickeisingboundstates}(b), (c) and (d) show the case of non-zero classical field, $\omega_x \neq 0$. 
Then, the Ising band \eqref{eq:isingdisp} is already open in the normal phase and we can understand the model as an impurity model, where the impurity role is played by the cavity. 

Before proceeding with the discussion, it is important to distinguish between the two Hamiltonians that we have presented so far, the original Dicke-Ising Hamiltonian defined in Eqs. \eqref{eq:Hzeeman}, \eqref{eq:Hising} and \eqref{eq:couplingising} and the mean-field effective Hamiltonian of Eq. \eqref{eq:HeffMFising}, which is useful for calculations of equilibrium and linear response properties, as we have shown. We have defined Bogoliubov fermions in Eq. \eqref{eq:Cxbogoliubovising} as the fermions that diagonalize the mean-field effective Hamiltonian. It is also possible however, to diagonalize the original Ising model \eqref{eq:Hising} and write the Dicke-Ising model in terms of the corresponding Bogoliubov fermions $\gamma_k$ and $\gamma_k^\dagger$ [written without $\sim$ in contrast with those of Eq. \eqref{eq:Cxbogoliubovising}]. 
For the following argument, we focus on the full Dicke-Ising Hamiltonian, such that the cavity appears explicitely.

The original collective coupling of the spins to the cavity translates into a momentum-dependent coupling of the Bogoliubov fermions that diagonalize the bare Ising model to the cavity [Cf. Eq. \eqref{eq:Cxbogoliubovising}]. 
Furthermore, it is possible to move from a momentum to a real-space representation with a Fourier transform, and express $C_x$ in terms of the domain-wall operators of the bare Ising model 
\begin{equation}
    C_x \propto i \sum_{mn} \eta_{m - n} \left(\gamma_m^\dagger \gamma_n^\dagger - \gamma_n \gamma_m\right) \,.
    \label{eq:Cxdomainwalls}
\end{equation}
Here, $\gamma^\dagger_n$ is a fermionic operator that upon acting on the bare Ising ground state creates a domain wall after the $n$-th spin \cite{prelovsek1981quantum}
\begin{equation}
    \gamma^\dagger_n = \frac{1}{\sqrt{N}} \sum_k \gamma_k^\dagger e^{-i k n} \,,
\end{equation}
and 
\begin{equation}
    \eta_j = \frac{1}{N} \sum_k \eta_k e^{i k j} \,,
\end{equation}
with
\begin{equation}
    \eta_k = 2 u_k v_k =  2J \sin k / \epsilon_k \,.
    \label{eq:etak}
\end{equation}

\begin{figure}
    \centering
    \includegraphics[width=\columnwidth]{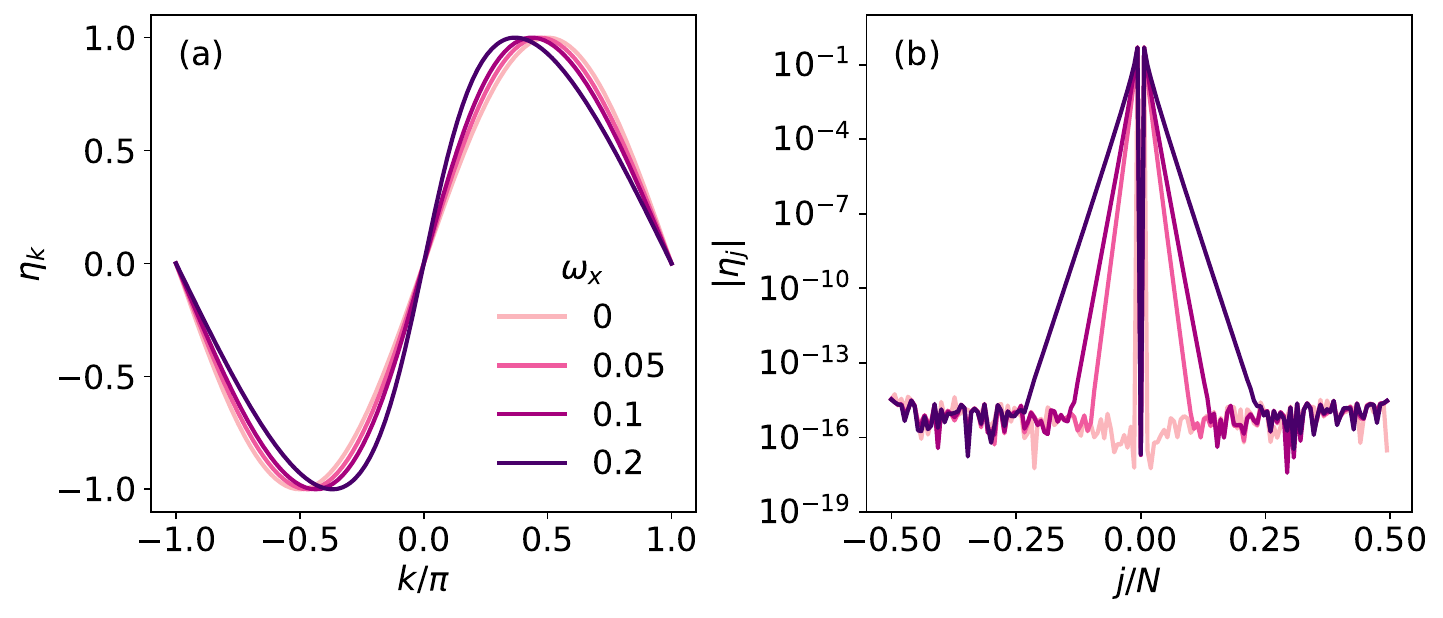}
    \caption{(a) Coupling of the Bogoliubov fermions, $\eta_k$, and (b) its Fourier transform, $\eta_j = \mathcal F\{\eta_k\}$, for different values of the classical transverse field, $\omega_x$. The Ising interaction is set to $4J = \Omega$ and the light-matter coupling to $\lambda = 0.2 \Omega$. A finite size of $N=150$ was used for the Fourier transform.}
    \label{fig:impurity}
\end{figure}

In Fig. \ref{fig:impurity} we show that $\eta_j$ is exponentially localized in real space, with maxima at $j = \pm 1$. Thus, from Eq. \eqref{eq:Cxdomainwalls}, we see that the cavity couples maximally to consecutive domain-walls $\gamma_m^\dagger \gamma_{m+1}^\dagger$, i.e. to single-spin domains, which is consistent with the fact that $C_x = \sum_j \sigma_j^x$ induces single spin flips. The coupling to wider domains $\gamma_m^\dagger \gamma_{m + w}^\dagger$ decreases exponentially with the width $w$ of the domain. 

The spin-boson model in waveguide QED can be described with a spin coupled to a single cavity of a cavity array. The spin acts as an impurity of the cavity array, inducing bound states in which photons are localized around the cavity to which the spin couples. Likewise, in the Dicke-Ising model the cavity couples only to narrow domains, which are a subset of the two-excitation Ising band. Thus, the cavity acts as an impurity of the two-excitation band, localizing spin-wave pairs around single-spin domains.

This interpretation is further supported by the fact that the equation for the poles of $D(\omega)$ can be shown to be
\begin{equation}
    F(\omega) = \omega^2 - \Omega^2 - 4 \lambda^2 \Omega \int_{- \pi}^\pi \frac{dk}{2 \pi} \tilde \eta_k^2 \frac{4 \tilde \epsilon_k}{\omega^2 - 4 \tilde \epsilon_k^2} = 0\,,
    \label{eq:polecondition}
\end{equation}
which can be compared with the equation for the eigenvalues of a discrete system coupled to a continuum with a finite bandwidth \cite{Bykov1975}, in our case the band of the Ising model given by \eqref{eq:isingdisp}. Note that $\tilde \eta_k$ and $\tilde \eta_j$ present only small quantitative differences with $\eta_k$ and $\eta_j$ displayed in Fig. \ref{fig:impurity} in the $z$FMN phase, where $m_x \ll 1$.

\subsection{Existence of Bound Polariton states}

By simple inspection, we observe in Figure \ref{fig:dickeisingboundstates} that 
$D(\omega)$ has poles outside the band given by $2 \tilde \epsilon_k$ in Eq. 
\eqref{eq:isingdisp}. We refer to these states as \emph{bound polariton 
states}. While we could simply use the term bound states, we retain the term 
polaritons since we are within the field of cavity QED materials. Additionally, 
this highlights their complementarity to the usual bound states in quantum 
optics, where matter localizes photons around an impurity. Here, it is the 
cavity that localizes spin excitations.

The possibility of bound states emerging is well known \cite{Bykov1975, John1990, gaveau1995limited}. These states belong to the discrete spectrum and are solutions to $F(\omega) = 0$ for $\omega$ outside the band. 
Much is known about bound states, particularly that their existence is primarily determined by the bandwidth, the coupling, and the position of the impurity \cite{shi2016bound, romanroche2020bound}, which in our case is the cavity.

Importantly, in this case, we are able to prove their existence as follows. 
First, we focus on the possibility of solutions below the lower band edge, 
given by $2 \tilde \epsilon_0$, \emph{i.e.}, $F(\omega < 2 \tilde \epsilon_0) = 0$. We first 
assume that $\Omega > 2 \tilde \epsilon_0$, the case depicted in the figure. Thus, 
the term $\omega^2 - \Omega^2 < 0$, and the integral in \eqref{eq:polecondition} 
is also negative. Therefore, for a solution to exist, the magnitude of the 
integral must be sufficiently large, which always occurs because the integral 
diverges as $\omega \to 2 \tilde \epsilon_0$.
On the other hand, if $\Omega < 2 \tilde \epsilon_0$, the condition for the existence of bound states is
\begin{equation}
 \Omega > 4 \lambda^2  \int_{-\pi}^\pi \frac{dk}{2 \pi} \frac{ \tilde \eta_k^2}{ \tilde \epsilon_k} = 0\,.
 \label{eq:poleconditionsimple}
\end{equation}
Similarly, the argument for bound states above the upper band, in this case $F(\omega > 2 \tilde \epsilon_\pi) = 0$, follows equivalent reasoning, proving or disproving their existence under the same conditions.

We consider it important to emphasize that these states emerge from a non-trivial 
system, which, despite being exactly solvable, is a strongly correlated model of 
matter non-perturbatively coupled to a cavity field. The fact that it can 
be solved highlights how useful the thermodynamic limit is in cavity QED 
materials when performing calculations at any value of the light-matter coupling.
Similar existence proofs should be obtained in other scenarios, such 
as intersubband polaritons \cite{debernardis2022magneticfieldinduced,cortese2019strong,cortese2020excitons} or lattice fermion models like the SSH model or 
similar \cite{dmytruk2022controlling, vlasiuk2023cavityinduced}. The existence of bound states in those cases can be discussed 
following the same procedure as here \cite{cortese2019strong}, and should depend on the band limits 
and the density of states of the matter coupled to the cavity near those band 
limits.

In Fig. \ref{fig:dickeisingboundstates}, we see how tuning the bandwidth with 
$\omega_x$ results in the detachment of bound states from the band edges. 
Furthermore, we understand their increased visibility (with respect to the 
band) when they appear, as it is well known that the contribution of the 
impurity (the cavity) is finite in the bound states. 
Additionally, the localized nature of bound states explains why the polariton 
formula \eqref{eq:2osci} accounts well for their energy.

\subsection{Finite-size effects}

\begin{figure}
    \centering
    \includegraphics[width=\columnwidth]{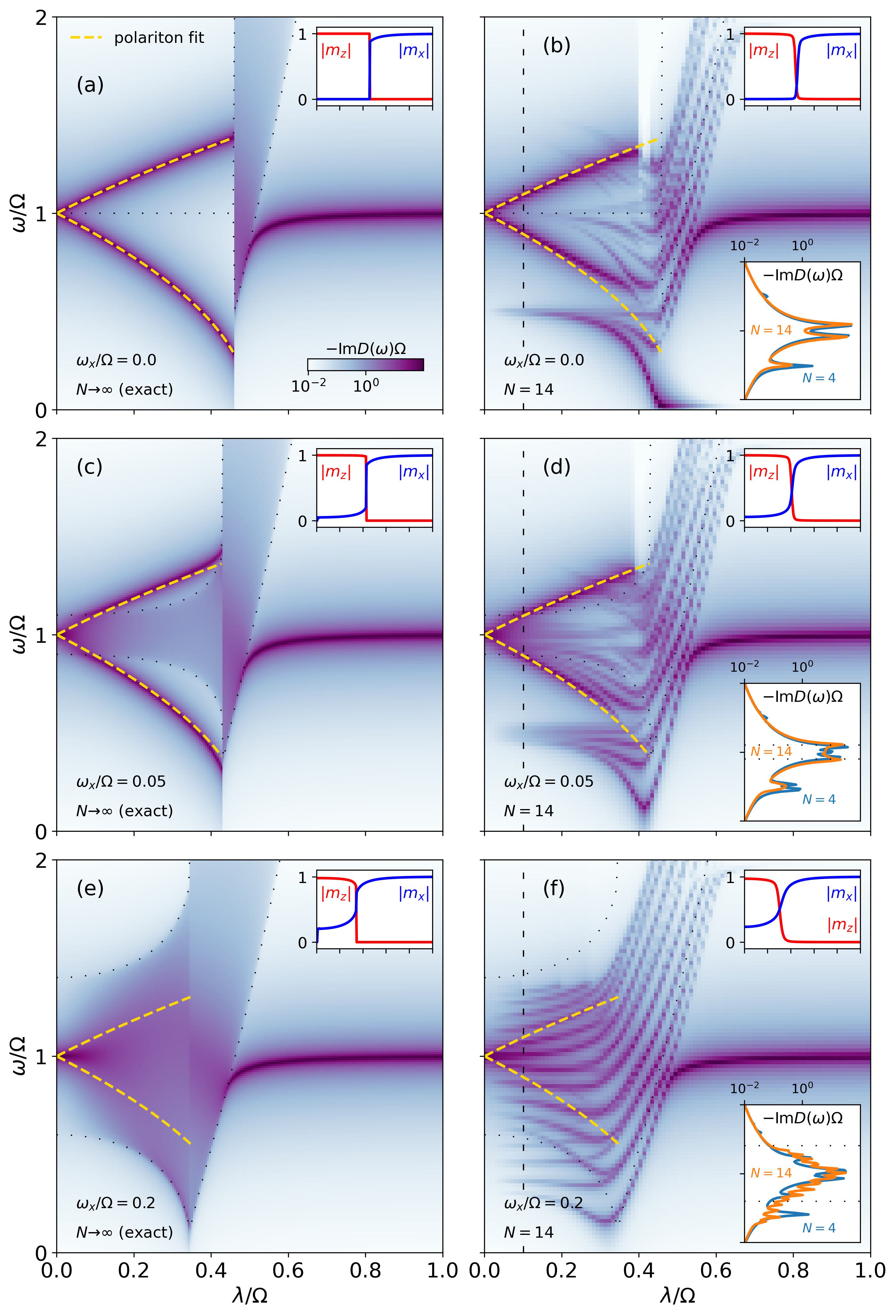}
    \caption{Cavity response, $D$, of the Dicke-Ising model as a function of the collective coupling, $\lambda$, computed analytically in the thermodynamic limit, $N\to\infty$, (left) and with exact diagonalization, $N=14$, (right) for different values of the classical field, $\omega_x$. 
    The yellow dashed lines correspond to a fit of the polaritons with a two-oscillator model [See Eq. \eqref{eq:2osci}]. 
    The top right insets show the magnetizations. 
    The bottom right inset in the right plots shows a vertical cut at the black dashed line for two finite sizes, $N=4$ and $N=14$. 
    The dotted lines mark the edges of the band of the mean-field effective Hamiltonian \eqref{eq:HeffMFising}. 
    The parameters are $\omega_x = 0$ and $4J = \Omega$. 
    In the exact-diagonalization results, the Fock basis for the photonic Hilbert space is truncated at 40 photons.}
    \label{fig:dickeisingexactdiag}
\end{figure}

Finally, we present a comparison between our analytical results of Sec. \ref{sec:solution} and finite-size exact-diagonalization results in Fig. \ref{fig:dickeisingexactdiag}. 
This allows us to discuss how quickly finite-size effects are washed out as we increase the system size.
We compare the analytical results in Figs. \ref{fig:dickeisingexactdiag}(a), (c) and (e), valid in the thermodynamic limit, with exact-diagonalization results for system sizes up to $N=14$ in Figs. \ref{fig:dickeisingexactdiag}(b), (d) and (f). 

Let us begin by comparing Figs. \ref{fig:dickeisingexactdiag} (a) and (b), which correspond to the case of vanishing classical field. 
In this case we observe the formation of polaritons in the normal phase and the opening of the two-excitation band and the hardening of the lower polariton in the superradiant phase in Fig. \ref{fig:dickeisingexactdiag}(a). 
The same features are observed in Fig. \ref{fig:dickeisingexactdiag}(b), although the two-excitation band is not fully formed and instead we can distinguish a collection of discrete levels. 
Additionally, there are some finite-size effects. 
Most prominently, there is a pole corresponding to the single-excitation band in the normal phase, at $\omega = 2J = \Omega/2$ for $\lambda \to 0$.
This is explained by noting that the coupling operator $C_x$ \eqref{eq:Cxbogoliubovising}, induces single-spin flips. 
To understand its effect, it is easier to reason in the limit of small light-matter coupling $\lambda$ and classical field $\omega_x$. 
Here, the spins are fully magnetized along $z$ in the ground state $|\uparrow \cdots \uparrow \rangle$ and the excitations are linear combinations of domain walls of the form $|\uparrow \cdots \uparrow \downarrow \cdots \downarrow\rangle$. When acting on the ground state, $C_x$ typically creates two contiguous domain walls, i.e. states with single-spin domains of the form $|\uparrow \cdots \uparrow \downarrow \uparrow \cdots \uparrow \rangle$, which belong to the two-excitation subspace. However, in an open finite chain, $C_x$ also connects the ground state with states of the form $|\uparrow \cdots \uparrow \downarrow\rangle$, which present a single domain wall and thus belong to the single-excitation subspace.
In the thermodynamic limit, these edge states represent a vanishing fraction of the single-excitation subspace and thus the visibility of the single excitation band in the photonic propagator becomes negligible as $ N \to \infty $.
This interpretation is confirmed by the bottom right inset in Fig. \ref{fig:dickeisingexactdiag}(b), which shows that the intensity of this pole decreases with size, unlike the poles corresponding to the polaritons. 
Figures \ref{fig:dickeisingexactdiag}(c), (d), (e) and (f) feature a finite classical field $\omega_x$ and thus a finite bandwidth in the normal phase. 
In Fig. \ref{fig:dickeisingexactdiag}(c) the narrow bandwidth allows the formation of bound polariton states with well defined energies below and above the band. 
This is validated in Fig. \ref{fig:dickeisingexactdiag}(d). 
Again, we observe additional features that we attribute to finite-size effects.
In particular the single-excitation band, at $\omega = 2J = \Omega/2$ for $\lambda \to 0$. 
Its visibility is shown to decrease with size in the bottom right inset of Fig. \ref{fig:dickeisingexactdiag}(d). 
In Fig. \ref{fig:dickeisingexactdiag}(e) the large bandwidth prevents the visibility of the bound polariton states. 
This is confirmed in Fig. \ref{fig:dickeisingexactdiag}(f) where we observe a collection of closely packed levels of equal visibility that are expected to form the band in the thermodynamic limit. 
The levels that fall well outside the would-be band are shown to be finite-size artifacts in the bottom right inset of Fig. \ref{fig:dickeisingexactdiag}(f).

\section{Conclusions}

In this work, we have studied the emergence of bound polaritons in cavity QED materials using the Dicke-Ising model, whose equilibrium and linear spectrum can be solved exactly in the thermodynamic limit.
We demonstrate the existence of bound polariton states, where the cavity-matter coupling leads to the localization of spin waves.  These results extend the polariton landscape and present new opportunities to control the dynamics of excitations in quantum materials through cavities.

Our work is based on the large $N$ theory for cavity QED materials \cite{romanroche2024cavity, romanroche2024linear}, 
which yields exact results. To explore how these effects persist in finite 
systems, we have performed exact-diagonalization calculations. We observe that 
even in modest system sizes, the described phenomenology is well reproduced, 
apart from the expected finite-size effects.
Finally, we have discussed how the physics described here is generalizable to 
other cases, such as intersubband polaritons or lattice fermion models like 
the SSH model coupled to cavities. Additionally, our work extends beyond 
light-matter systems, including to phonon-polaritons 
\cite{pantazopoulos2024unconventional} or magnon-spin coupling
\cite{kim2024observation}.

\section*{Acknowledgements}
We acknowledge discussions with Yuto Ashida. Despite the enormous impediments in spending the budget, the authors must acknowledge funding from the grant
TED2021-131447B-C21 funded by MCIN/AEI/10.13039/501100011033 and the EU
‘NextGenerationEU’/PRTR. 
We also acknowledge grant CEX2023-001286-S financed by MICIU/AEI /10.13039/501100011033, the Gobierno de Aragón (Grant E09-17R Q-MAD), Quantum Spain and the CSIC Quantum
Technologies Platform PTI-001. J. R-R acknowledges support from the Ministry of Universities of the Spanish Government through the grant FPU2020-07231.

\appendix

\section{Microscopic theory for magnetic cavity QED}
\label{app:microscopic}

Here we consider a microscopic derivation of the light-matter Hamiltonian. A complementary perspective can be obtained by employing macroscopic QED theory \cite{martinez2024general}.

For simplicity, we consider a system of $N$ neutral single-electron atoms, each with a non-dynamical nucleus with charge $e$ at position $\bs R_j$ and a dynamical electron with charge $-e$, bound to the nucleus, at position $\bs R_j + \bs r_j$.
The derivation can be readily extended to multi-electron atoms or molecules.
Note that $\bs R_j$ is a classical variable and $\bs r_j$ is an operator. 
The charge density of the system reads
\begin{equation}
    \rho(\bs r) = -e \sum_j^N\delta^3 (\bs r - \bs R_j - \bs r_j) + e \sum_j^N\delta^3 (\bs r - \bs R_j) \,,
\end{equation}
and one can define polarization
\begin{equation}
    \bs P(\bs r) = -e \sum_j^N\bs r_j \int_0^1 ds \delta^3 (\bs r - \bs R_j - s \bs r_j) \,
\end{equation}
and magnetization
\begin{equation}
    \bs M(\bs r) = -e \sum_j^N\bs r_j \times \dot{\bs r}_j \int_0^1 ds s \delta^3(\bs r - \bs R_j - s \bs r_j) \,
\end{equation}
fields that satisfy $\nabla \cdot \bs P = - \rho$ and $\nabla \times \bs M = \bs j - \delta_t \bs P$, with the current density $\bs j(\bs r) = -e \sum_j^N\dot{\bs r}_j \delta^3 (\bs r - \bs R_j - \bs r_j)$.
Here $\delta^3(\bs r)$ is the three-dimensional Dirac delta function.

We start from the Hamiltonian in the Coulomb gauge
\begin{equation}
\begin{multlined}
    H = \sum_j^N\frac{\left(\bs p_j + e \bs A(\bs R_j + \bs r_j) \right)^2}{2m} + V_{\rm C} + V 
    \\
    + H_{\rm em} + \frac{g_{\rm e} \mu_{\rm B}}{2} \sum_j^N\bs \sigma_j \cdot \bs B(\bs R_j + \bs r_j) \,,
\end{multlined}
\end{equation}
In the Coulomb gauge, the redundancy in the description of the dynamical variables is eliminated by constraining the vector potential to be transverse, i.e. $\nabla \cdot \bs A = 0$. 
The quantized vector potential then reads
\begin{equation}
    \bs A(\bs r) = \sum_{\kappa} A_{\bs k} \bs u_{\kappa}(\bs r) a_{\kappa}  + {\rm h.c.} \,, 
    \label{eq:vectorpotential}    
\end{equation}
where $\kappa$ is a four-vector containing the polarization index $\sigma = 1,2$ and the wavevector $\bs k$: $\kappa \equiv \{\bs k, \sigma\}$.
The mode amplitude is $A_{\bs k} = \sqrt{1/(2\epsilon_0 V \omega_{\bs k})}$. 
The $V$ and $\epsilon_0$ are respectively the cavity mode volume and the dielectric constant.
We also introduce the bosonic annihilation and creation operators of the $\kappa$-th mode: $a_{\kappa}$, $a_{\kappa}^\dagger$, which obey the canonical commutation relations $[a_{\kappa}, a_{\kappa'}^\dagger] = \delta_{\kappa \kappa'}$. 
To maintain as much generality as possible, we have refrained from using a particular spatial dependence for the vector potential. 
For a specific model, the geometry of the cavity will determine the spatial quantization of the wavevector $\bs k$ and in turn, the functional form of the mode functions $\bs u_{\kappa}(\bs r)$ \cite{kakazu1994}. 
In any case, the following properties hold for the mode functions: $\bs u_{-\bs k, \sigma}(\bs r) = \bs u_{\bs k, \sigma}^*(\bs r)$, $\bs u_{\bs k, \sigma}(\bs r) \cdot \bs k = 0$, $\int_V dV \bs u^*_{\kappa}(\bs r) \cdot \bs u_{\kappa'}(\bs r) = \delta_{\kappa \kappa'}$. 
Note that in Eq. \eqref{eq:vectorpotential} the mode functions have been promoted to mode operators, since they depend on the position operator of each particle. 
By definition, $\bs B = \nabla \times \bs A$, so
\begin{equation}
    \bs B(\bs r) = \sum_{\kappa} B_{\bs k} \bs u_{\perp, \kappa} (\bs r)  a_{\kappa}  + {\rm h.c.} \,,
    \label{eq:magneticfield}
\end{equation}
where we have defined the transverse mode functions $\bs u_{\perp, \kappa} (\bs r) = |\bs k|^{-1} \nabla \times \bs u_{\kappa} (\bs r)$ and $B_{\bs k} = A_{\bs k} |\bs k| = A_{\bs k} \omega_{\bs k} / c$.
Finally, we can express $ H_{\rm em}$ as 
\begin{equation}
     H_{\rm em} = \sum_{\kappa}  \omega_{\bs k} a_{\kappa}^\dagger a_{\kappa} \,.
\end{equation}

To move to the multipolar gauge we will use the Power-Zienau-Woolley (PZW) transformation, defined as
\begin{equation}
    U = e^{-i \int d^3 r \bs P(\bs r) \cdot \bs A(\bs r)}.
\end{equation}
The resulting Hamiltonian in the multipolar gauge reads \cite{steck2007quantum}
\begin{equation}
\begin{multlined}
    H = \sum_j^N\frac{\left(\bs p_j - e \bs r_j \times \int_0^1 ds s \bs B(\bs R_j + s \bs r_j) \right)^2}{2m} \\+ V_{\rm C} + V + \sum_\kappa \omega_{\bs k} a_\kappa^\dagger a_\kappa \\
    - i \sum_\kappa \omega_{\bs k} A_{\bs k} \int d^3 r \bs P(\bs r) \cdot \left( \bs u_\kappa(\bs r) a_\kappa - {\rm h.c.}\right) \\
    + \sum_\kappa \omega_{\bs k} \left| A_{\bs k} \int d^3 r \bs P(\bs r) \cdot \bs u_\kappa(\bs r) \right|^2  \\
    + \frac{g_{\rm e} \mu_{\rm B}}{2} \sum_j^N\bs \sigma_j \cdot \bs B(\bs R_j + \bs r_j)\,.
\end{multlined}
\end{equation}
We can now perform the long-wavelength approximation for the electromagnetic field. 
Instead of taking the crudest approximation, we will perform a multipolar expansion of the polarization and the magnetic field up to the electric quadrupole and magnetic dipole terms. 
For the polarization we approximate
\begin{equation}
    \delta^3(\bs r - \bs R_j - s \bs r_j) \approx \delta^3(\bs r - \bs R_j) -s(\bs r_j \nabla)\delta^3(\bs r - \bs R_j) \,
\end{equation}
and subsequently
\begin{equation}
    \bs P(\bs r) = -e \sum_j^N\bs r_j \delta^3(\bs r - \bs R_j) + \frac{1}{2} e \sum_j^N \bs r_j (\bs r_j \cdot \nabla) \delta^3(\bs r - \bs R_j) \,.
\end{equation}
With this
\begin{equation}
\begin{multlined}
    \int d^3r \bs P(\bs r) \cdot \bs u_\kappa (\bs r) = -e \sum_j^N\bs r_j \cdot \bs u_\kappa(\bs R_j) 
    \\+ \frac{1}{2} e \sum_j^N\bs r_j \cdot \left[(\bs r_j \cdot \nabla) \bs u_\kappa(\bs r)\right]_{\bs r = \bs R_j} \,.
\end{multlined}
\end{equation}
Similarly, we approximate the magnetic field as $\bs B(\bs R_j + \bs r_j) \approx \bs B(\bs R_j)$. 
The resulting Hamiltonian reads
\begin{equation}
\begin{multlined}
    H = H_\rmm - \sum_j^N\bs m_j \cdot \bs B(\bs R_j) + \frac{1}{8m} \sum_j^N\left(\bs d_j \times \bs B(\bs R_j)\right)^2  \\
    + \sum_k\omega_{\bs k} a_\kappa^\dagger a_\kappa -i \sum_\kappa \omega_{\bs k} A_{\bs k} \sum_j^N\bs d_j \cdot (\bs u_\kappa(\bs R_j) a_\kappa - {\rm h.c.}) \\
    + i \sum_\kappa \frac{\omega_{\bs k}^2}{c} A_{\bs k}  \sum_j^N \left( {\rm Tr}(Q_j U_\kappa(\bs R_j) a_\kappa - {\rm h.c.}\right) \\
    +\sum_\kappa \omega_{\bs k} \left|A_{\bs k} \sum_j^N\bs d_j \cdot \bs u_\kappa (\bs R_j) - A_{\bs k} \frac{\omega_{\bs k}}{c} \sum_j^N{\rm Tr}Q_j U_\kappa(\bs R_j) \right|^2 \,,\\
\end{multlined}
\label{eq:Hmultipolar}
\end{equation}
with
\begin{align}
    &\bs d_j = -e \bs r_j \,, \\
    &Q_{j, \alpha \beta} = -\frac{1}{2} e \left(r_{j, \alpha} r_{j, \beta} - \frac{\bs r_j^2}{3} \delta_{\alpha \beta}\right) \,, \\
    &\bs m_j = -\mu_{\rm B} \left( \bs r_j \times \bs p_j + \frac{g_{\rm e}}{2} \bs \sigma_j \right)
\end{align}
the electric dipole, electric quadrupole and magnetic dipole operators, respectively. 
To simplify notation, we have defined $U_\kappa(\bs R_j) = |\bs k|^{-1} J_{\bs u_\kappa}(\bs R_j)$ with $J_{\bs u_\kappa}$ the Jacobian matrix of $\bs u_\kappa$ at $\bs R_j$.
This simplified Hamiltonian \eqref{eq:Hmultipolar} reveals the main forms of atom-light interaction. 
The second term corresponds to the Zeeman coupling between the total magnetic moment of the electron and the magnetic field. 
The third term, coupling the electric dipole to the magnetic field, is known as the diamagnetic term. 
We will disregard it, as it is negligible compared to the others for small per-particle couplings \cite{steck2007quantum}.
Note that we could have identified 
\begin{equation}
    \bs E^\perp(\bs r) = -\partial_t \bs A(\bs r) =  i \sum_\kappa \omega_{\bs k} A_{\bs k}  \left(\bs u_\kappa(\bs r) a_\kappa - {\rm h.c.}\right) \,,
\end{equation}
in the fifth and sixth terms, which couple the electric dipole and quadrupole to the transverse electric field, respectively \cite{stokes2022implications}. 
The last term corresponds to a self-interaction of the polarization field and is typically referred to as the $P^2$ term. 
With this, and defining the constants
\begin{align}
    & g_{\bs k}^\rmm = \frac{\omega_{\rm p}}{\sqrt{N}} \sqrt{\frac{\omega_{\bs k}}{8mc^2}} \label{eq:gm} \, \\
    & g_{\bs k}^{\rm d} = \frac{\omega_{\bs k}}{\sqrt{N}} \sqrt{\frac{\omega_{\rm p}}{\omega_{\bs k}}} \,, \label{eq:gd} \\
    &  g_{\bs k}^{\rm q} = \frac{\omega_{\bs k}}{\sqrt{N}} \sqrt{\frac{2 \omega_{\bs k}}{m c^2}} \label{eq:gq} \,,
\end{align}
we can put together the interaction terms to define electric
\begin{equation}
\begin{multlined}
    C_\kappa^{\rm e} = -i \sum_j^N\Biggl(\sqrt{\frac{m \omega_{\rm p}}{2}} 
    \bs d_j \cdot \bs u_\kappa (\bs R_j) \\
    - \frac{g_{\bs k}^{\rm q}}{g_{\bs k}^{\rm d}} \frac{m \omega_{\rm p}}{2} {\rm Tr}(U_\kappa(\bs R_j) Q_j) \Biggr) \,, \\
\end{multlined}
\end{equation}
and magnetic coupling operators
\begin{equation}
    C_{\kappa}^\rmm = -\sum_j^N\frac{1}{\mu_{\rm B}} \bs m_j \cdot \bs u_{\perp, \kappa}(\bs R_j) \,,
\end{equation}
to finally write the Hamiltonian as
\begin{equation}
\begin{multlined}
    H = H_\rmm + \sum_{\kappa}  \Omega_{\bs k}  a_{\kappa}^\dagger a_{\kappa} + \sum_{\kappa} \frac{g_{\bs k}^{\rm d}}{\Omega_{\bs k}} \left| C_\kappa^{\rm e}\right|^2 \\
    + \sum_\kappa \left(g_{\bs k}^{\rm d} C_\kappa^{\rm e} + g_{\bs k}^\rmm C_\kappa^{\rm m} \right) a_{\kappa}  + {\rm h.c.} \,,    
\end{multlined}
\label{eq:linearlmHemitters}
\end{equation}
with $\Omega_{\bs k} = \omega_{\bs k}$.
The $P^2$ term prevents the electronic coupling to the cavity from modifying the ground-state properties with respect to the bare matter model, which already includes electrostatic Coulomb interactions \cite{romanroche2022effective, romanroche2024cavity, lenk2022collective}.
This is a no-go theorem akin to those derived in the Coulomb gauge from the $A^2$ term \cite{andolina2019cavity, andolina2020theory, andolina2021nogo, rokaj2018lightmatter, schafer2020relevance, lamberto2024quantum}.
The $A^2$ and $P^2$ terms ensure gauge invariance in these models \cite{diStefano2019resolution, garziano2020gauge, savasta2021gauge, dmytruk2021gauge, li2020electromagnetic}.
This motivates the consideration of purely magnetic materials, with negligible polarizability and thus negligible electrostatic dipole-dipole interactions and coupling of the electric dipole to the cavity's electric field \cite{romanroche2021photon}.
These materials interact with the cavity via the Zeeman coupling of the magnetic dipole and the cavity's magnetic field.
Under these assumptions, and considering for simplicity a single uniform cavity mode, we arrive at Eq. \eqref{eq:Hzeeman}.

\section{Computing $\tilde \chi_{xx, 0}$ for the Ising model}
\label{app:responseising}

The spectral decomposition of $\tilde \chi_{xx, 0}$ reads
\begin{equation}
    \tilde \chi_{xx, 0} = -\frac{1}{N} \sum_n |\langle n | C_x | 0 \rangle|^2 \frac{2 (E_n - E_0)}{\omega_+^2 - (E_n - E_0)^2} \,,
\end{equation}
where $|n\rangle$ is the eigenstate of $H_{\rm{eff}}^{\rm{MF}}$ \eqref{eq:HeffMFising} with eigenvalue $E_n$. From Eq. \eqref{eq:Cxbogoliubovising} we see that 
\begin{equation}
    \langle n | C_x | 0 \rangle = \delta_{n0} \left(N + \sum_k v_k^2\right) -2 i \sum_k u_k v_k \langle n | \gamma_k^\dagger \gamma_{-k}^\dagger | 0 \rangle \,.
\end{equation}
With this and using Eq. \eqref{eq:etak} we find
\begin{equation}
    \tilde \chi_{xx, 0} = -\frac{16 J^2}{N} \sum_k \frac{\sin^2 k}{\epsilon_k (\omega_+^2 - 4 \epsilon_k^2)}
\end{equation}
and in the thermodynamic limit, $\lim_{N \to \infty} N^{-1} \sum_k f_k = \int_{-\pi}^{\pi} dk / (2 \pi) f_k$, we obtain Eq. \eqref{eq:chi0ising}.

\bibliography{main.bib}

\end{document}